\documentclass[aps,pra,noshowpacs,twocolumn,preprintnumbers,superscriptaddress,floatfix]{revtex4}

\usepackage{amsmath,amssymb}
\usepackage{multirow}
\usepackage{color}
\usepackage{hyperref,graphicx}
\usepackage{epsfig}
\usepackage{times}
\usepackage{pdfpages}

\begin{document}
\title{A broadband DLCZ quantum memory in room-temperature atoms}

\author{Jian-Peng Dou
$^\dagger$}
\affiliation{State Key Laboratory of Advanced Optical Communication Systems and Networks, School of Physics and Astronomy, Shanghai Jiao Tong University, Shanghai 200240, China}
\affiliation{Synergetic Innovation Center of Quantum Information and Quantum Physics, University of Science and Technology of China, Hefei, Anhui 230026, China}
\author{Ai-Lin Yang
$^\dagger$}
\affiliation{State Key Laboratory of Advanced Optical Communication Systems and Networks, School of Physics and Astronomy, Shanghai Jiao Tong University, Shanghai 200240, China}
\affiliation{Synergetic Innovation Center of Quantum Information and Quantum Physics, University of Science and Technology of China, Hefei, Anhui 230026, China}
\author{Mu-Yan Du}
\affiliation{State Key Laboratory of Advanced Optical Communication Systems and Networks, School of Physics and Astronomy, Shanghai Jiao Tong University, Shanghai 200240, China}
\author{Di Lao}
\affiliation{State Key Laboratory of Advanced Optical Communication Systems and Networks, School of Physics and Astronomy, Shanghai Jiao Tong University, Shanghai 200240, China}
\author{Jun Gao}
\affiliation{State Key Laboratory of Advanced Optical Communication Systems and Networks, School of Physics and Astronomy, Shanghai Jiao Tong University, Shanghai 200240, China}
\affiliation{Synergetic Innovation Center of Quantum Information and Quantum Physics, University of Science and Technology of China, Hefei, Anhui 230026, China}
\author{Lu-Feng Qiao}
\affiliation{State Key Laboratory of Advanced Optical Communication Systems and Networks, School of Physics and Astronomy, Shanghai Jiao Tong University, Shanghai 200240, China}
\affiliation{Synergetic Innovation Center of Quantum Information and Quantum Physics, University of Science and Technology of China, Hefei, Anhui 230026, China}
\author{Hang Li}
\affiliation{State Key Laboratory of Advanced Optical Communication Systems and Networks, School of Physics and Astronomy, Shanghai Jiao Tong University, Shanghai 200240, China}
\affiliation{Synergetic Innovation Center of Quantum Information and Quantum Physics, University of Science and Technology of China, Hefei, Anhui 230026, China}
\author{Xiao-Ling Pang}
\affiliation{State Key Laboratory of Advanced Optical Communication Systems and Networks, School of Physics and Astronomy, Shanghai Jiao Tong University, Shanghai 200240, China}
\affiliation{Synergetic Innovation Center of Quantum Information and Quantum Physics, University of Science and Technology of China, Hefei, Anhui 230026, China}
\author{Zhen Feng}
\affiliation{State Key Laboratory of Advanced Optical Communication Systems and Networks, School of Physics and Astronomy, Shanghai Jiao Tong University, Shanghai 200240, China}
\affiliation{Synergetic Innovation Center of Quantum Information and Quantum Physics, University of Science and Technology of China, Hefei, Anhui 230026, China}
\author{Hao Tang}
\affiliation{State Key Laboratory of Advanced Optical Communication Systems and Networks, School of Physics and Astronomy, Shanghai Jiao Tong University, Shanghai 200240, China}
\affiliation{Synergetic Innovation Center of Quantum Information and Quantum Physics, University of Science and Technology of China, Hefei, Anhui 230026, China}
\author{Xian-Min Jin}
\thanks{xianmin.jin@sjtu.edu.cn\\
$^\dagger$These authors contributed equally.}
\affiliation{State Key Laboratory of Advanced Optical Communication Systems and Networks, School of Physics and Astronomy, Shanghai Jiao Tong University, Shanghai 200240, China}
\affiliation{Synergetic Innovation Center of Quantum Information and Quantum Physics, University of Science and Technology of China, Hefei, Anhui 230026, China}
\date{\today}

\maketitle
\textbf{Quantum memory capable of stopping flying photons and storing their quantum coherence is essential for scalable quantum technologies. A room-temperature broadband quantum memory will enable the implementation of large-scale quantum systems for real-life applications. Due to either intrinsic high noises or short lifetime, it is still challenging to find a room-temperature broadband quantum memory beyond conceptual demonstration. Here, we present a far-off-resonance Duan-Lukin-Cirac-Zoller (FORD) protocol and demonstrate the broadband quantum memory in room-temperature atoms. We observe a low unconditional noise level of ${\bf 10^{-4}}$ and a cross-correlation up to 28. A strong violation of Cauchy-Schwarz inequality indicates high-fidelity generation and preservation of non-classical correlation. Furthermore, the achieved cross-correlation in room-temperature atoms exceeds the key boundary of 6 above which quantum correlation is able to violate Bell's inequality. Our results open up the door to an entirely new realm of memory-enabled quantum applications at ambient conditions.}

\section*{Introduction}
\noindent Quantum technologies, incorporating quantum mechanics into communication, information processing and metrology, promise spectacular quantum enhanced advantages beyond what can be done classically \cite{OBrien2009}. However, quantum states are very fragile and easily get lost to the environment, meanwhile their generation and quantum operations are mostly probabilistic. These problems make it exponentially hard to build long-distance quantum channel for quantum communications \cite{Jin2010,Gisin2007} and large quantum systems for quantum computing \cite{Ladd2010,Guzik2012}. Quantum memory \cite{Lvovsky2009} allows quantum states to be stored and retrieved in a programmable fashion, therefore providing an elegant solution to the probabilistic nature and associated limitation by coordinating asynchronous events \cite{Briegel1998,Duan2001,Nunn2013}. 

Enormous advances in quantum memory have been made by developing various photon storage protocols and their physical implementations, such as electromagnetically induced transparency (EIT) \cite{Chaneliere2005,Eisaman2005,Zhang2011}, Duan-Lukin-Cirac-Zoller (DLCZ) memory \cite{Duan2001,Kuzmich2003,Chrapkiewicz2017}, off-resonant Faraday interaction \cite{Julsgaard2004}, controlled reversible inhomogeneous broadening \cite{Moiseev2001,Alexander2006}, atomic frequency combs \cite{Afzelius2009} and Raman memory \cite{Reim2011,Ding2015}. In order to have quantum memory practicable for efficient synchronisation and physical scalability, considerable efforts have been dedicated to meet key features known as high efficiency, low noise level, large time bandwidth product (lifetime divided by pulse duration) and operating at room temperature \cite{Lvovsky2009}.

It has proven very difficult to satisfy all the requirements simultaneously. Especially, in the regime of large bandwidth and room temperature, noise and/or decoherence become dominant, therefore memories can not work in quantum regime \cite{Manz2007,Michelberger2015} or only in short time \cite{Lee2011,England2015,Kaczmarek2018,Finkelstein2018}. At room temperature, EIT and near off-resonance Raman memory have a collision-induced fluorescence noise which can not be filtered out because of being identical with the signal photons \cite{Manz2007}. By applying larger detuning, far off-resonance Raman is found to be able to well eliminate the fluorescence noise and posses a large storage bandwidth. Unfortunately, noise rising from spontaneous Raman scattering process becomes dominant, which is intrinsic and proportional to the detuning \cite{Michelberger2015}.

Here we present FORD protocol where we exploit spontaneous Raman scattering process to generate and store an excitation rather than taking it as noise. We demonstrate a genuine broadband quantum memory that can simultaneously meet aforementioned key features (see Supplementary Figure 1, Supplementary Table 1, Note 1 and Note 2 in Supplementary Information). We have observed an unconditional noise level (the noise counts divided by the trial number of read process without write process) of $10^{-4}$ and a cross-correlation between heralding Stokes photon and retrieved anti-Stokes photon up to 28. A violation of Cauchy-Schwarz inequality \cite{Clauser1974} up to 20 standard deviations indicates high-fidelity generation and preservation of non-classical correlation in room-temperature atoms. Furthermore, the achieved cross-correlation in room-temperature atoms exceeds the key boundary of 6 above which quantum correlation are able to violate Bell's inequality \cite{de Riedmatten2006,Bao2012}. A time bandwidth product of 700 can be promptly employed to build large-scale quantum networks. 

\begin{figure*}[ht!]
\centering
\includegraphics[width=0.65 \textwidth]{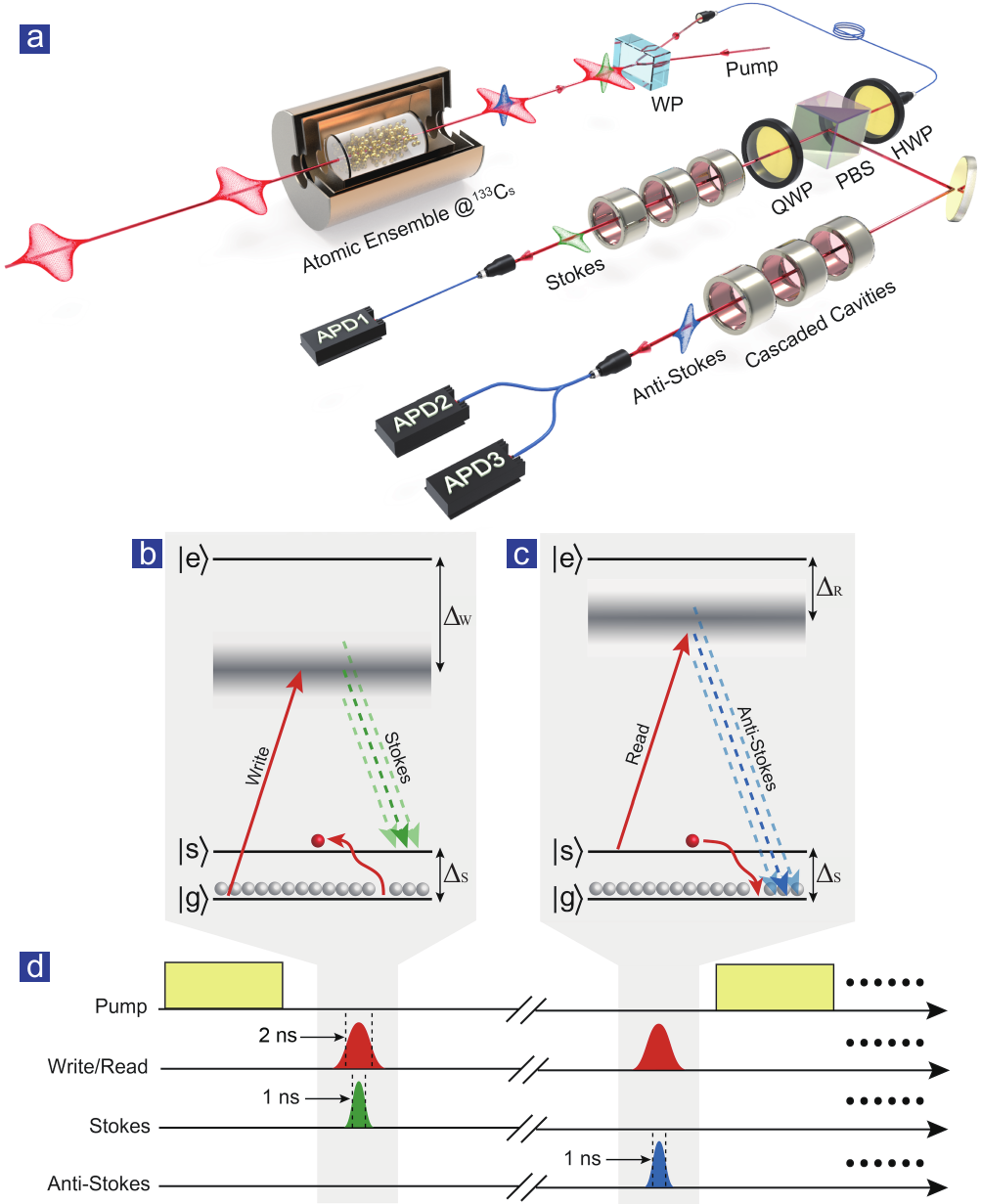}
\caption{\textbf{Experimental setup and FORD scheme}. \textbf{a.} The caesium cell is packed in a three-layer magnetic shielding and is heated up to 61.3\,$^{\circ}$C. The write and read pulses (red envelopes) are generated with a programmed time delay and are both prepared in horizontal polarization (see Methods for more details). The created Stokes photons (green envelope) and retrieved anti-Stokes photons (blue envelope) are both in vertical polarization. A Wollaston prism (WP) is employed as polarization filter to separate the output photons from the write and read pulses. The two sets of cascaded cavities serve as strong spectrum filters and contribute an extinction ratio up to $10^{7}$. The colours are a guide to the eye. HWP: half-wave plate, QWP: quarter wave plate, PBS: polarization beam splitter. \textbf{b.} The write process of FORD quantum memory. The blurred gray belt denotes broadband virtual excited state induced by write pulse. The green dash lines represent wide transition band of a Stokes photon. $\Delta_{\rm S}=9.2\,$\rm GHz is the ground state hyperfine splitting of caesium. \textbf{c.} The read process of FORD quantum memory. The blue dash lines represent wide transition band of an anti-Stokes photon. \textbf{d.} The time sequence for generation, storage and retrieval of nonclassical correlation. Ellipsis implies repeated sequences afterward.}
\label{Figure 1}
\end{figure*}

\begin{figure*}
\centering
\includegraphics[width=0.75\textwidth]{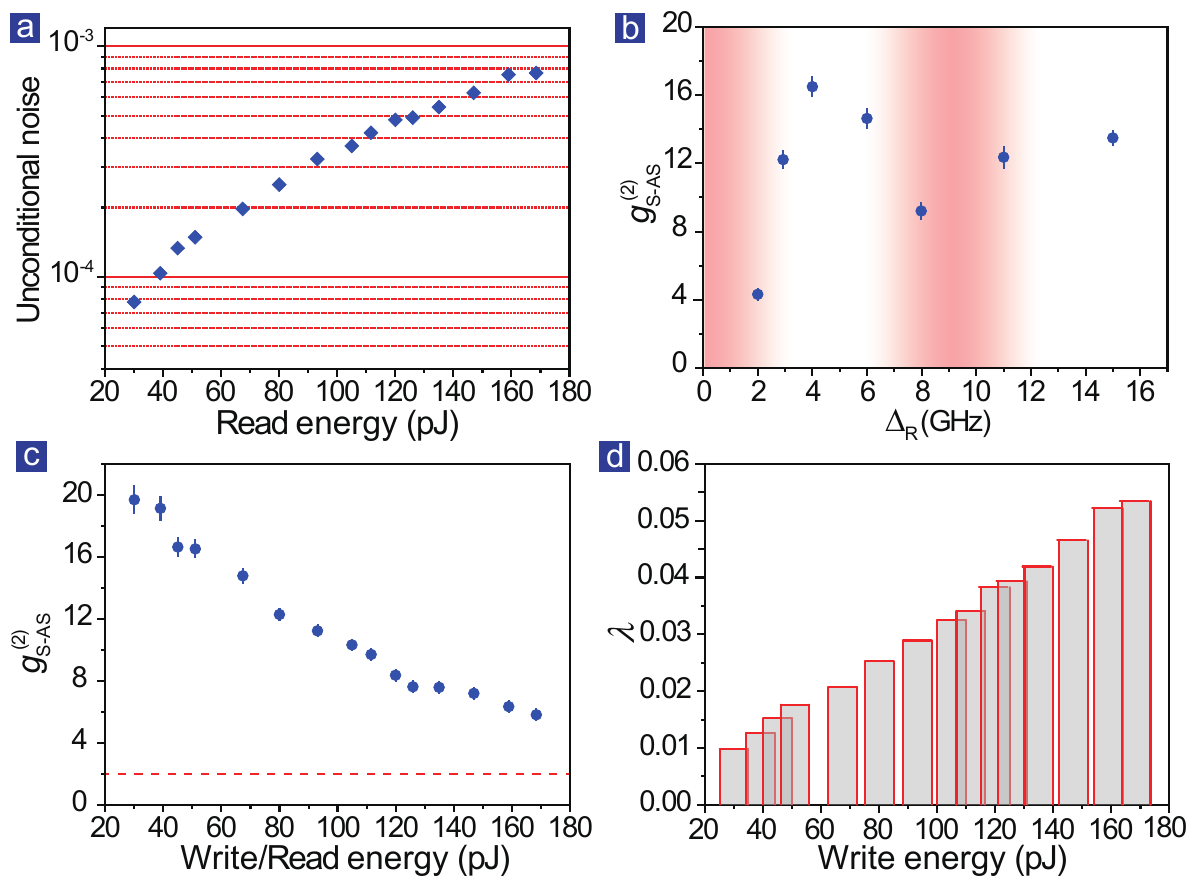}
\caption{\textbf{Experimental results on the noise level and nonclassicality}. \textbf{a.} The measured unconditional noise level as a function of the read pulse energy. \textbf{b.} The measurement of cross-correlation in a large detuning from near to far off-resonance region, $\Delta_{\rm R}$ from 2\,GHz to 15\,GHz, correspondingly $\Delta_{\rm W}=\Delta_{\rm R}+9.2\,\rm GHz$. \textbf{c.} Influence of the write/read pulse energy. $g_{\rm S\text{-}AS}^{(2)}$ above the boundary (the dash line) implies non-classical correlation. Error bars are derived on the basis of the Poisson distribution of single photon detectors. \textbf{d.} The relation between excitation probability $\lambda$ and the write pulse energy. }
\label{Figure 2}
\end{figure*}

\section*{\bf Results}
\noindent{\bf FORD quantum memory scheme.} As is shown in Figure 1a, in contrast to ``mapping in and out" of external photons in other quantum memory protocols, DLCZ memory creates one collective excitation directly inside the atoms by a classical write pulse via spontaneous Raman process, meanwhile emits a Stokes photon which can herald the successful storage with intrinsic unit efficiency \cite{Duan2001,Afzelius2015}. With a programmable delay, a read pulse can retrieve the stored excitation as anti-Stokes photon with a process similar to Raman memory protocol, which has also been demonstrated with the potential to approach a unit efficiency using a strong enough read pulse \cite{Reim2011}. 

By combining far off-resonance atomic configuration and standard DLCZ process, an excited virtual energy level near the two-photon resonance can be created by the strong write/read pulse, see Figure\,1b,\,1c\,and\,1d. The linewidth of the excited virtual energy level is proportional to the intensity of the write/read laser and the effective optical depth of atomic ensemble. In our experiment, we use broadband write and read pulse with a pulse duration of 2\,ns and detune them to a very far off-resonance region. The detuning in linear frequency $\Delta_{\rm W}$ and $\Delta_{\rm R}$ are larger than 13.2\,GHz and 4\,GHz respectively, which is about one order higher than narrowband DLCZ quantum memory protocol. We adopt ceasium atoms $^{\rm 133}$Cs to realize large optical depth relying on its high vapour pressure. With 75-mm-long cell and 10\,Torr Ne buffer gas, an optical depth larger than 1000 is obtained at a temperature of 61.3\,$^{\circ}$C. The storage bandwidth is expected to be near GHz level and its central frequency is tunable by detuning the write and read pulse either simultaneously or separately.

A simplified three-level $\Lambda$-type configuration is illustrated in Figure 1b and 1c. The lower two energy states $\left | g \right \rangle \left( 6S_{1/2}, F=3\right)$ and $\left | s \right \rangle\left( 6S_{1/2}, F=4\right)$ are the hyperfine ground states of caesium, and the higher energy state $\left | e \right \rangle  \left( 6P_{3/2}, F'=2, 3, 4, 5 \right)$ is the excited state. Initially, a pumping laser resonant with the transition of $ 6S_{1/2}, F=4 \rightarrow 6P_{3/2}, F'= 5 $ prepare all the atoms into the state $\left | g \right \rangle$ as
\begin{equation}\label{eq01}
\left | g_1 g_2 \cdots  g_N \right \rangle ,
\end{equation}
where $N$ is the total number of atoms that participate in interaction. After the initial state is prepared, a strong write pulse with a detuning $\Delta_{\rm W}$ creates a single excitation  among billions of atoms meanwhile induces a broadband Stokes photon via spontaneous Raman scattering \cite{Duan2001,Duan2002} (see Methods). In this process, appropriate energy of the write pulse is adopted to ensure that only a single Stokes photon is emitted per attempt and high-order emissions are negligible. Then the collective excitation state can be expressed as \cite{Sangouard2011}
\begin{equation}\label{eq02}
\frac{1}{\sqrt{N}}\sum_{n=1}^{N} e^{i\left( {\bf k}_{\rm W}-{\bf k}_{\rm S}\right){\bf r}_n} \left | g_1 g_2 \cdots g_{n-1} s_n g_{n+1} \cdots g_N \right \rangle ,
\end{equation}
where ${\bf k}_{\rm W}$ is the wave vector of the write pulse, ${\bf k}_{\rm S}$ is the wave vector of the Stokes photon, and ${\bf r}_{n}$ is the position of the $n$th atom. After a programmable delay, a strong read pulse with a detuning $\Delta_{\rm R}$ and wave vector ${\bf k}_{\rm R}$ illuminates the caesium cell and transform the collective excitation into a broadband anti-Stokes photon whose wave vector is denoted by ${\bf k}_{\rm AS}$.

\begin{figure*}
\centering
\includegraphics[width=1.98\columnwidth]{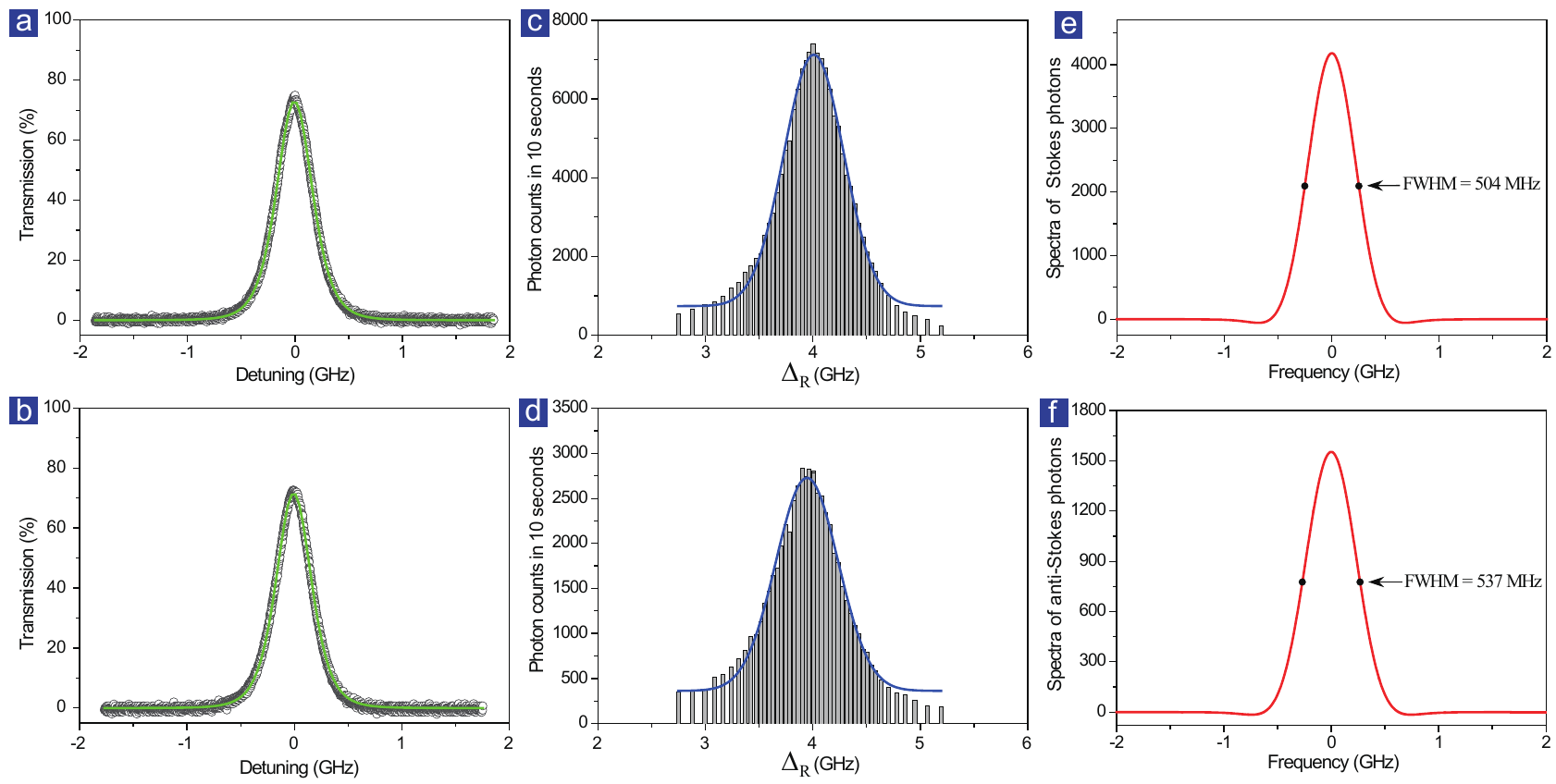}
\caption{\textbf{Experimental results of convolution-based bandwidth measurement}. We measure the total transmission windows of cascaded cavities for Stokes (\textbf{a}) and anti-Stokes photons (\textbf{b}). We then detune the write/read light to get another transmission spectra for Stokes (\textbf{c}) and anti-Stokes photons (\textbf{d}). We deduce the frequency spectra of Stokes (\textbf{e}) and anti-Stokes photons (\textbf{f}) by using convolution theorem and Fourier transform.}
\label{Figure 3}
\end{figure*}

The write and read pulse are coaxial, which implies ${\bf k}_{\rm W}={\bf k}_{\rm R}$, in order to maximise the spin wave lifetime of atomic excitation \cite{Zhao2009}. It has been theoretically demonstrated that the Stokes photons are mainly inside a small cone around the direction of the write pulse \cite{Duan2002}. According to the phase-matching condition ${\bf k}_{\rm W}+{\bf k}_{\rm R}={\bf k}_{\rm S}+{\bf k}_{\rm AS}$ , we infer that the Stokes and anti-Stokes photons are approximately coaxial as well. As shown in Figure 1a, both Stokes and anti-Stokes photons are polarized orthogonally to the write/read pulse \cite{Nunn2008} and therefore can be separated from the strong addressing light with a high-extinction Wollaston polariser (see Notes 3 and 4 in Supplementary Information for more details). Two sets of homemade cascaded cavities composed of three Fabry-Perot cavities are tuned resonant with Stokes and anti-Stokes photons respectively. Each cavity has a transmission rate about 90$\%$ and contributes an extinction ratio more than 500. Together with a polarising beam splitter and a quarter wave plate, we realise a dichroic-mirror-like functioning for an extremely small frequency difference of 9.2 GHz (see Methods for more details). A standard Hanbury-Brown and Twiss interferometer composed of three avalanched photodiodes and a 50:50 fibre beamsplitter are employed to perform photon statistics and correlation detection.\\

\begin{figure}
\centering
\includegraphics[width=0.98\columnwidth]{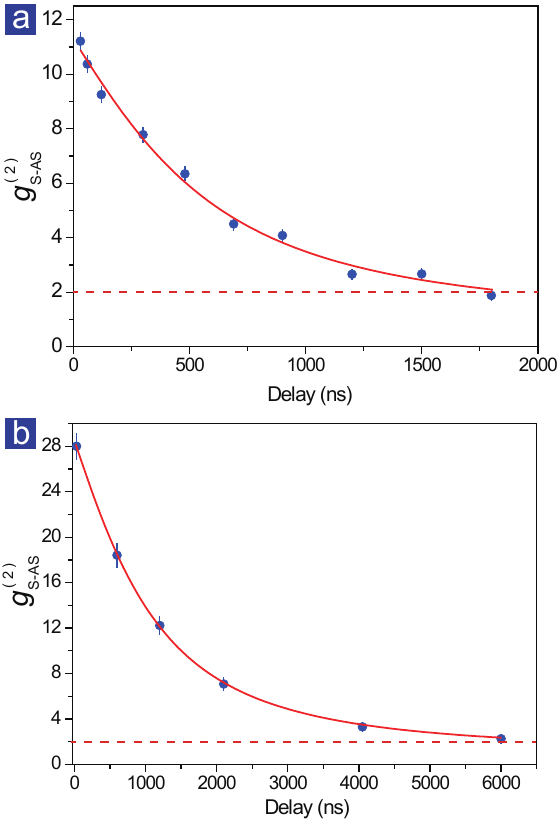}
\caption{\textbf{Lifetime measurement.} \textbf{a.} The measured cross-correlation as a function of storage time. The data is obtained under the condition of a write/read beam waist of 90\,$\mu$m, pulse energy about 100\,pJ and the detuning $\Delta_{\rm R}=4$\,GHz. \textbf{b.} The observed improvement of lifetime and cross-correlation by expanding the beam waist to 240 \,$\mu$m. Error bars are derived on the basis of the Poisson distribution of single photon detectors. The red curve is a theoretical fit of the form $g_{\rm S\text{-}AS}^{(2)}=1+C/(1+At^2+Bt)$.} 
\label{Figure 4}
\end{figure}

\noindent{\bf Observation of broadband quantum memory in room-temperature atoms.} Unconditional noise is a key parameter that can be used to benchmark the noise level of optical memory and whether it can work at quantum regime. We measure unconditional noise by retrieving anti-Stokes photons in absence of the write pulse. As a result of the novel protocol and experimental configuration, we observe a  low unconditional noise level of $10^{-4}$. We obtain a value of $7.79(5)\times10^{-5}$ when we address atoms with the read pulse energy of 30\,pJ, see Figure\,2a. A distinct manifestation of such a low unconditional noise level presents a strong nonclassical correlation between the heralding Stokes photon and the retrieved atomic excitation anti-Stokes photon. It turns out to be a high cross-correlation $g_{S\text{-}AS}^{(2)}$ and a violation of Cauchy-Schwarz inequality $(g_{\rm S\text{-}AS}^{(2)})^{2}\leq g_{\rm S\text{-}S}^{(2)}\cdot g_{\rm AS\text{-}AS}^{(2)}$ \cite{Clauser1974}. We obtain an auto-correlation of $g_{\rm S\text{-}S}^{(2)}=1.97\pm0.13$ ($g_{\rm AS\text{-}AS}^{(2)}=1.87\pm0.20$ ) for Stokes photons (anti-Stokes photons) and cross-correlation of $7.83\pm0.18$, with the detuning $\Delta_{\rm R}$ at 4\,GHz, the write/read pulse energy 129\,pJ and the delay time 30\,ns. The Cauchy-Schwarz inequality is violated up to 20 standard deviations (see Note 5 in Supplementary Information for the calculation formula), which clearly indicates a high-fidelity generation and preservation of non-classical correlation in our quantum memory.

It is important but technically challenging to investigate storage performance depending on detuning of addressing light. For every detuning data point in Figure\,2b, we realise this by detuning and locking the write/read frequency far away from the transition line of caesium, and also initialising the resonance for all cascaded cavities. We have made measurement of cross-correlation in a large detuning range $\Delta_{\rm R}$ from 2\,GHz to 15\,GHz. For broadband optical memory at room temperature, the performance ranging from near to far off resonance is identified. Our results show a ``sweet spot" but inconsistent with previous results in narrowband Raman memory experiments which identify 1.3\,GHz as the optimal detuning \cite{Bashkansky2012}. Apart from the detuning around 0\,GHz or 9.2\,GHz where the leakage of fluorescence through cascaded cavities is not negligible, FORD quantum memory shows the ability of well working at quantum regime in a wide spectrum (see Note 2 in Supplementary Information). We also made the measurement of cross-correlation and created probability of atomic excitation by scanning the write/read pulse energy, which reveals a higher cross-correlation up to 20 at lower excitation probability, see Figure\,2c\,and\,2d.

In order to obtain the bandwidth of our quantum memory, we develop a convolution-based approach. We measure the total transmission spectra of cascaded cavities by scanning a narrowband classical light, see Figure\,3a\,and\,3b. We then count Stokes and anti-Stokes photons while scanning the detuning of the write/read light, shown in Figure\,3c\,and\,3d. By using convolution theorem and Fourier transform (see Methods), we can deduce the frequency spectra of Stokes and anti-Stokes photons shown in Figure\,3e\,and\,3f. With the write/read pulse energy of 96\,pJ, the measured bandwidth of Stokes and anti-Stokes photons are 504\,MHz and 537\,MHz respectively, which is about two order higher than the bandwidth obtained in cold atoms \cite{Chaneliere2005,Eisaman2005,Zhang2011}.

The time bandwidth product is a key figure of merit of quantum memory endowed with the capacity of synchronisation. This parameter sets the limit of the times that we can synchronise within the lifetime of stored correlations. We measure $g_{\rm S\text{-}AS}^{(2)} $ as a function of storage time, see Figure\,4a. The data is fitted with the form $g_{\rm S\text{-}AS}^{(2)}=1+C/(1+At^2+Bt)$ where $At^2$ results from atomic random motion \cite{Zhao2009}, and $Bt$ is a correction term that reflects the effect of the background noise of write and read light  (see Supplementary Table 2 and Note 6 in Supplementary Information for more details). The lifetime is defined with the cross-correlation dropping to 1/e, which is found to reach 800\,ns, see Figure 4a. The fitting function reveals that at this stage the lifetime is limited by the nonzero background of address light and the motion induced loss of atoms. Through enlarging the mode field of both the write/read pulse, we achieve a prolonged lifetime to 1400\,ns and obtain a time bandwidth product up to 700, see Figure 4b. We also observe a higher cross-correlation of up to 28 and being higher than 2 until time delay is 6000\,ns.

\section*{Discussion}
The distinct feature of the FORD scheme is that we exploit a spontaneous Raman scattering process to generate and store an excitation rather than taking it as noise at the write process. It means the write process is completely free of noise. The read process of the FORD scheme is the same as the far off-resonance Raman scheme suffering from the four-wave mixing noise. Another distinct feature of the FORD scheme is that we apply differential coupling strength via detuning for the write and read process. The spontaneous Raman scattering can be well suppressed in the read process because the read field strongly couples with atoms and retrieves the stored excitation back to photon with a probability as high as possible. We identify the performance ranging from the near to far off-resonance. Our results show a ``sweet spot" of detuning that is different from the value adopted in previous experiments. At the optimal detuning, the four-wave mixing noise is well suppressed. We therefore are able to observe a low unconditional noise level of $10^{-4}$.

The off-resonance Raman process has been proven to be able to reach unit retrieval efficiency \cite{Reim2011}. Our current retrieval efficiency conditional on registering a Stokes photon is around 10\%. We attribute the gap to the noise, the limited read pulse energy and the mode mismatch of coupling Stokes and anti-Stokes photons. It should be noticed that in our experiment a single pulse laser acts as both the write and read field. The stored photon has the chance to be retrieved instantaneously during the write process, which may set an upper limit for storage efficiency. However, the correlated coupling can be suppressed by employing two control lasers so that the momentum, frequency, bandwidth and intensity of the write and read pulse can be tuned individually, which makes it possible to optimize one without disturbing the other. Further improvements include increasing the read pulse energy \cite{Reim2011}, shaping the read pulse \cite{Novikova2008} and cavity enhancement \cite{Nunn2016}.

Although the time bandwidth product has already been very high, a longer absolute lifetime means fewer quantum repeater nodes for a given quantum communication distance \cite{Briegel1998,Sangouard2011}. The dominant factor of decoherence mechanism in our current setup is random motion induced loss, which may be solved by using a small-diameter cell to keep atoms always staying in the interaction region. Meanwhile anti-relaxation coating should be adopted to keep atomic polarization during its collision with the cell. The lifetime of Zeeman populations and coherences in excess of 60 seconds has been reported in \cite{Balabas2010}, making it potential for realizing longer storage lifetime (see Note 7 in Supplementary Information).

Apart from quantum repeaters, more importantly, quantum memory enabled synchornization is crucial to build large-scale multi-photon and quantum entanglement states for quantum computing, quantum simulation and quantum metrology. Our quantum memory is more straightforward for experimental implementation (i.e. physically more scalable) and therefore suitable for such applications. 

In summary, we have demonstrated a broadband DLCZ quantum memory in room-temperature atoms. Low unconditional noise level, strong nonclassicality preserved among heralding photon and stored excitation, and large time bandwidth product and ability of operating at room temperature make FORD quantum memory promising for future scalable quantum technologies \cite{OBrien2009,Whiting2017}, and promptly applicable in building large-scale quantum networks \cite{Kimble2008}.

\section*{Methods}
\noindent{\bf Programmable generation of high-intensity write/read pulse.} 
Far off-resonance DLCZ quantum memory requires broadband and high-intensity write/read pulse. We cannot generate it directly from a high-power continuous laser since the required peak power exceeds the threshold of fast Electro Optic Modulator (EOM). A customised commercial Ti:sapphire laser may provide required bandwidth and intensity. However, fixed periodic generation of pulse means incapability to write and read on-demand, resulting in inadequate exploitation of time bandwidth product of quantum memory. We develop a system to generate high-intensity pulse with tunable central frequency, bandwidth, and more importantly, generation time. An external cavity diode laser (ECDL) locked to the transition $6S_{1/2},F= 4\rightarrow6S_{3/2},F'=4\,{\rm co}\,5$ line of caesium serves as reference called MASTER. Another distributed Bragg reflector (DBR) laser called SLAVE are locked to MASTER in an arbitrarily set frequency difference by comparing their frequency on a Fabry-Perot etalon. A fast EOM trigged by electronic pulses from field-programmable gate array (FPGA) is used to chop SLAVE to short pulses. The generated pulses are fed into a homemade tapered amplifier (TA) to boost their power by 17 dB. In order to eliminate the spontaneous emission from the TA, we employ a ruled diffraction grating to spread beam out and spatially pick the stimulated radiation with irises. In our experiment, full width at half maximum (FWHM) of the write/read pulse is about 2 \,ns, measured by a fast photodiode whose rise time is tens of picoseconds.\\

\noindent{\bf The excitation is shared among billions of atoms.}
Many theories and experiments have confirmed the collective enhancement of signal-to-noise ratio during write process and collective interference effect during read process. These collective effects verify that the excitation is shared among a large number of atoms rather than a single atom. The well-known references have been included in the main text \cite{Duan2001, Duan2002, Sangouard2011}.

The atomic number density $n$ can be roughly estimated by $n = \frac{{4\pi d}}{{\lambda ^2 L}}$ , where optical depth $d$ is about $5000$, wavelength $\lambda \approx 852\,{\rm nm}$, cell length $L = 75\,{\rm mm}$ . The estimated number density  is about $1.15 \times 10^{18} /{\rm m^3} $. We can estimate the total atomic number among the interaction region (the beam waist of laser is on the order of $100\,\mu {\rm{m}}$ ) as 
\begin{equation}
\begin{split}
N &=  n \times \pi  \times \left( {{\rm{Beam waist}}} \right)^2  \times {\rm{cell length}} \\
  &=  1.15 \times 10^{18} /{\rm{m}}^3  \times \pi  \times \left( {100\mu {\rm{m}}} \right)^2  \times 75{\rm{mm}}\\
& \approx 2.7 \times 10^9 .
\end{split}
\end{equation}
So the excitation is shared among billions of atoms.

In each write process, there is a small probability $\lambda$ to emit a Stokes photon denoted by $\left| {1_{\rm p} } \right\rangle$ corresponding to a collective excitation denoted by $\left| {1_{\rm a} } \right\rangle$ stored in the atomic ensemble. The total wave function can be written as \cite{Duan2001}
\begin{equation}
\left| {\Phi _{\rm ap} } \right\rangle  = \left| {0_{\rm a} 0_{\rm p} } \right\rangle  + \sqrt \lambda  \left| {1_{\rm a} 1_{\rm p} } \right\rangle  + \lambda \left| {2_{\rm a} 2_{\rm p} } \right\rangle  + O\left( {\lambda ^{3/2} } \right)
\end{equation}
$ O\left( {\lambda ^{3/2} } \right)$ represents the high order terms with more excitations whose probabilities are equal to or smaller than ${\lambda ^{3/2} }$.
The excitation probability $\lambda$  shown in Figure\,{2}d is controlled by the energy of the write pulse. We can see that the excitation probability is linear to the energy of the write pulse, which is consistent with the statement of cold atoms based experiments, for example, the reference \cite{Chou2004}.\\

\noindent{\bf Cascaded cavities.} Collinear configuration of writing and reading offers longer spin wave lifetime while makeing single photons very hard to be filtered out of strong addressing light. We develop two sets of cascaded cavities composed of three Fabry-Perot cavities tuned resonant with Stokes and anti Stokes respectively. Every cavity is a monolithic plank-convex glass. Both sides are coated to give an overall transmission window FWHM about 380\,MHz. Careful alignment is necessary to optimise mode matching between cavity and incident light, which determines the performance in terms of transmission and extinction ratio. The transmission frequency can be tuned by setting the temperature of the cavity. We use an active feedback system to set the temperature and also lock it within $\pm$\,3\,mK. For each cavity, we obtain a transmission rate of more than 90$\%$ and an extinction ratio of more than 500. With three cavities together we have a transmission rate over 70$\%$ and total extinction ratio up to $10^{7}$.\\

\noindent{\bf How the frequency filter works.} The frequency filter composed of a polarization beam splitter, a quarter wave plate and two sets of cascaded cavities, which are resonant with Stokes and anti-Stokes photons respectively. After the Wollaston polarizer, Stokes and anti-Stokes photons are coupled into a single mode fiber. Then the photons with horizontal polarization pass through the polarization beam splitter and encounter the filter.  Firstly the photons encounter the Stokes-resonant cavity, the Stokes photons pass through the Stokes-resonant cavity, but the anti-Stokes photons are reflected by the front surface of the Stokes-resonant cavity. Then the reflected anti-Stokes photons pass through the quarter wave plate again. Note that, double pass of a quarter wave plate is equivalent to a half wave plate. The anti-Stokes photons therefore flip its polarization to vertical and reflected by the polarization beam splitter, and finally  pass through the anti-Stokes-resonant cavity. We can see that the Stokes photons and anti-Stokes photons pass through different set of cavities respectively. There is no 50\% loss of photons.   

The transmission windows of cavities shown in Figure\,3a and 3b are measured with a weak classical light from a distributed Bragg reflector laser. Single photons are not the best choice for measuring transmission windows of cavities since the fluctuations of single photons and ambient noise will bring considerably difficulty to adjust and optimize the cavities. On the other hand, high intensive light is also not suitable since the cavities will be warmed by the intensive light and the transmission windows of cavities will shift away from their correct frequency position.\\

\noindent{\bf  Convolution-based bandwidth measurement.} As shown in Figure\,3a\,and\,3b, the total transmission windows of the cascaded cavities are fitted by using
\begin{equation}\label{eq03}
 \begin{aligned}
T\left(f\right)= &\frac{T_0}{1+A\sin^2\left[d\left(f-f_0\right)\right]} \cdot\frac{1}{1+B\sin^2\left[h\left(f-f_0\right)\right]}\cdot \\
&\frac{1}{1+C\sin^2\left[g\left(f-f_0\right)\right]}
 \end{aligned}
\end{equation}
where $T_0$, $A$, $B$, $C$, $d$, $h$, $g$ and $f_0$ are fitting parameters. Detuning refers to frequency difference $f-f_0$. The grey circles are experimental data and the green lines are fitting curves. Figure\,3c\,and\,3d are fitted by using
\begin{equation}\label{eq04}
U\left(f\right)=a\exp{\left[\frac{-2\left(f-b\right)^2}{d^2}\right]}+U_0 ,
\end{equation}
where $a$, $b$, $d$ and $U_0$ are fitting parameters. The grey columns are experimental data and the blue lines are fitting curves. It is reasonable to consider $U_0$ as noise rather than signal photons, so it is $U\left(f\right)-U_0$ rather than $U\left(f\right)$ that describes real convolutions. If function $S\left(f\right)$ is assumed to be the frequency spectra of Stokes photons or anti-Stokes photons, then
\begin{equation}\label{eq05}
U\left(f\right)-U_0=T\left(f\right)*S\left(f\right),
\end{equation}
Applying convolution theorem, we achieve
\begin{equation}\label{eq06}
F\{S\left(f\right)\}=\frac{F\{U\left(f\right)-U_0\}}{ F\{T\left(f\right)\}} ,
\end{equation}
where $F\{S\left(f\right)\}$ is the Fourier transform of $S\left(f\right)$, similarly for $F\{U\left(f\right)-U_0\}$ and $F\{T\left(f\right)\}$.
By applying inverse Fourier transform to $F\{S\left(f\right)\}$, we derive out the spectrum $S\left(f\right)$ of Stokes photons or anti-Stokes photons. The spectra are shown in Figure\,3e\,and\,3f.\\

\bigskip
\textbf{Data availability.} The data that support the findings of this study are available from the corresponding author on reasonable request.

\section*{Acknowledgements}
The authors thank Jian-Wei Pan and Hui-Jun Li for helpful discussions. This work was supported by National Key R\&D Program of China (2017YFA0303700); National Natural Science Foundation of China (NSFC) (11374211, 61734005, 11690033); Shanghai Municipal Education Commission (SMEC)(16SG09, 2017-01-07-00-02-E00049); Science and Technology Commission of Shanghai Municipality (STCSM) (15QA1402200, 16JC1400405,17JC1400403). X.-M.J. acknowledges support from the National Young 1000 Talents Plan.

\section*{Author contributions} X.-M.J. conceived the project. J.-P.D, A.-L.Y. and X.-M.J. designed the experiment. J.-P.D., A.-L.Y., M.-Y.D., D.L., J.G., L.-F.Q., H.L., X.-L.P., Z.F., H.T. performed the experiment. X.-M.J. and J.-P.D. analysed the data and wrote the paper. 

\section*{Competing Interests} The authors declare that they have no competing interests.

\clearpage
\newpage

\onecolumngrid
\section*{\large Supplemental Information: A broadband DLCZ quantum memory in room-temperature atoms}
\setcounter{figure}{0}
\setcounter{table}{0}
\setcounter{equation}{0}
\renewcommand{\figurename}{Supplementary Figure}
\renewcommand{\tablename}{Supplementary Table}

\renewcommand{\thetable}{\arabic{table}}
\renewcommand{\theequation}{{S}\arabic{equation}}

\bigskip
\section*{\large Supplementary Note 1: The advances of quantum memory in atomic ensemble}
\begin{figure}[b]
\includegraphics[width=0.96\textwidth]{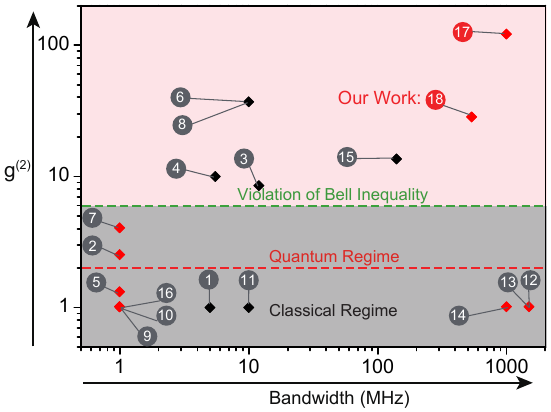}
\caption{Milestone works of quantum memory towards broadband and quantum regime in atomic ensemble. The quantum memory experiments in cold atoms were shown in black diamond. The quantum memory experiments in room-temperature atoms were shown in red diamond. }
\label{Supplementary Figure 1}
\end {figure}
\begin{table}[]
	\centering
\includegraphics[width=0.96\textwidth]{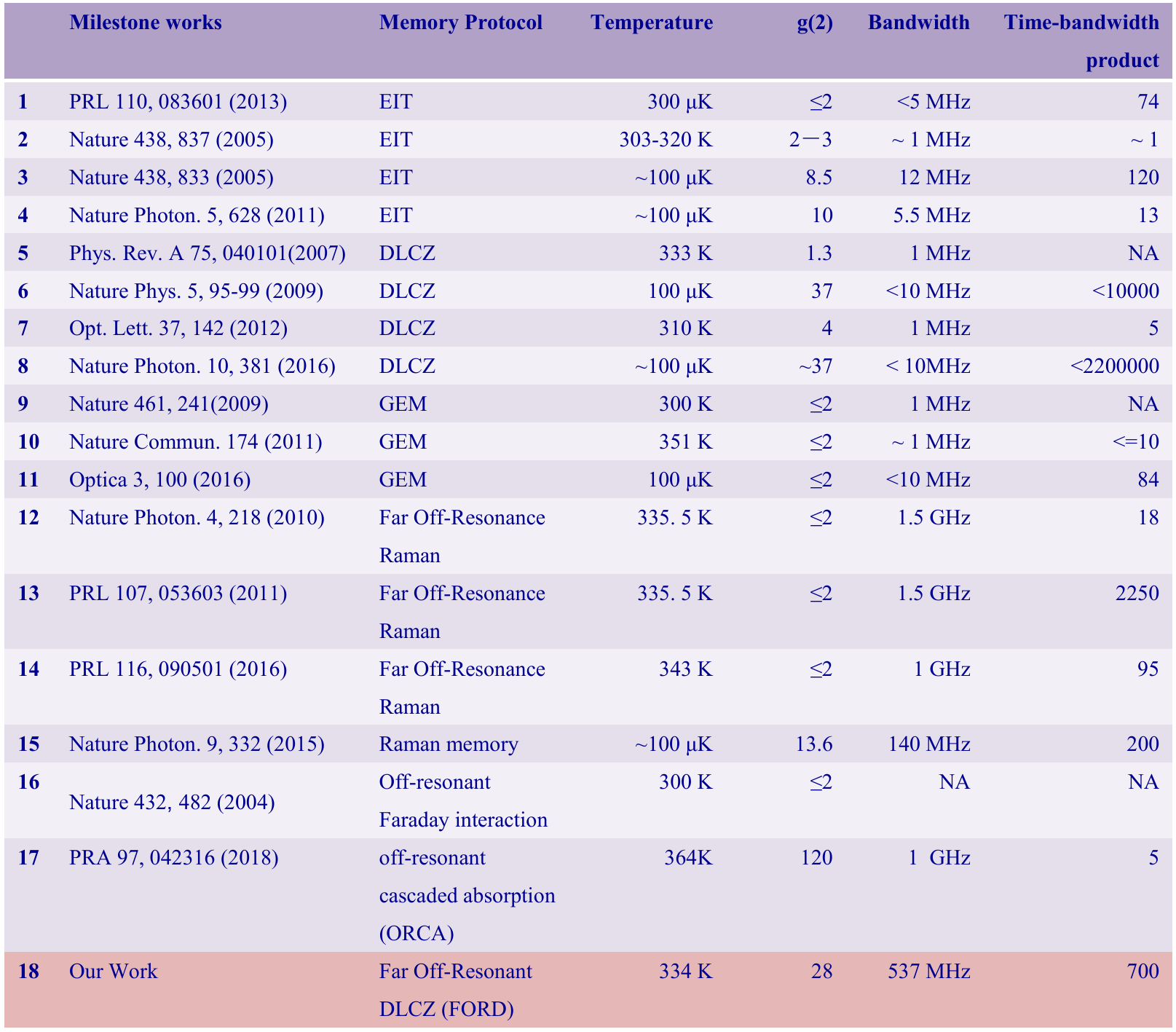}
\caption{Milestone works on quantum memory in atomic ensemble and key figures of merit.}
\label{Table1}
\end {table}
Enormous milestone works have been made to achieve quantum memory practicable for scalable quantum information. Apart from quantum repeaters, more importantly, quantum memory enabled synchornisation is crucial to build large-scale multi-photon and quantum entanglement states for quantum computing, quantum simulation and quantum metrology. 
Key features include high efficiency, low noise level, large time bandwidth product and operating at room temperature, most importantly, operating in quantum regime. Here we make very careful literature searching and list representative milestone works as well as their main achieved results (visually shown in Supplementary Figure\,1 and more details in Supplementary Table 1). We can see that the nonclassicality obtained in most previous milestone experiments in room-temperature atoms is below or around the quantum-classical boundary (see the red dashed in Supplementary Figure\,1), and more importantly below the boundary that can violate Bell's inequality and therefore for genuine quantum applications. 
The violation of Bell's inequality is characterized by $S > 2$ , where $S$ may be expressed as $S \approx 2\sqrt 2 V$ with the visibility $V \approx {{\left( {g_{\rm S - AS}^{(2)}  - 1} \right)} \mathord{\left/ {\vphantom {{\left( {g_{\rm S - AS}^{(2)}  - 1} \right)} {\left( {g_{\rm S - AS}^{(2)}  + 1} \right)}}} \right. \kern-\nulldelimiterspace} {\left( {g_{\rm S - AS}^{(2)}  + 1} \right)}}$ that define the boundary as $g_{\rm S - AS}^{(2)} = 6$ (see the green dashed in Supplementary Figure\,1)\cite{de Riedmatten2006,Bao2012}. That is why we have never seen any single-photon quantum teleportation, quantum swapping, quantum repeater and quantum internet with room-temperature quantum memory. Our work and the work 17 were finished almost at the same time, and both for the first time broke the long-standing bottleneck of realizing broadband quantum memories at room-temperature atoms. The time-bandwidth product achieved in our work is about 140 times higher, which therefore genuinely opened up the door to practical memory-enabled quantum applications at ambient conditions.

\section*{\large  Supplementary Note 2: The far off-resonant DLCZ (FORD) protocol against noise} 
The main problem that ever prevented us from realizing room-temperature quantum memory is noise. In 2005, Lukin's group from Harvard University observed nonclassicality with near resonance DLCZ protocol in hot atoms. The obtained nonclassicality is only around the classical boundary, and the experiments were performed in a very low atomic density (extremely low memory efficiency while no much collision-induced fluorescence noise) \cite{Eisaman2005}. After that, many groups have repeated the protocol but failed to realize a reasonably useful room-temperature quantum memory. In 2007, Jian-wei Pan's group in Heidelberg University experimentally confirmed a no-go conclusion that resonant or near resonant DLCZ memory in hot atoms can not work in quantum regime due to the collision-induced fluorescence noise \cite{Manz2007}. In our work, we present and experimentally demonstrate a far off-resonance DLCZ protocol to eliminate the influence of the collision-induced fluorescence noise.
 
The noise level is affected by many factors, including pressure of buffer gas, collision fluorescence, filtering performance, beam waist, temperature of atoms associated with collisions and Doppler broadening, etc. The Doppler effect and atom-atom collision induce a fluorescence broadening about 1\,GHz. The detuning we used in our experiment should be larger than 1 GHz. For a $\Lambda$-type configuration, the fluorescence noises mainly distribute around 0\,GHz and 9.2\,GHz respectively. We use two light red bands to illustrate the fluorescence noises, which is shown in Figure\,2b. The optimal detuning is around 4 GHz which is far away from the two fluorescence noise ranges. In a simple model, the larger detuning we applied, the lower noise level we will get. However, in large detuning region, we have to use stronger write and read pulse to address the atoms, which induces additional noise depending on the performance of polarization and spectral filters. Thus it is very hard to model and fit experimental results because for different detuning the collision-induced fluorescence noise is different, and the write and read pulse energy (also affect noise level) is also different in order to have a same excitation probability. We can see that the noise highly depends on the specific experimental setting. Of course, with back and forth modifications one can make a curve to predict or fit experimental results, but this is not a rigorous way to treat our data in Figure\,2b. For broadband optical memory at room temperature, we for the first time identify the performance ranging from the near to far off-resonance. Our results show a "sweet spot" of detuning that is different from the value adopted in previous experiments.

\section*{\large  Supplementary Note 3: Remarks of co-propagation scheme}
The co-propagation scheme possesses many advantages:

1. In the co-propagation scheme, the maximum wavelength of spin wave can be achieved, which means a spin-wave dephasing induced decoherence can be minimized for higher limit of lifetime \cite{Zhao2009}. 

2. It has been theoretically demonstrated that the Stokes photons more likely emit forward inside a small cone along the direction of the write pulse \cite{Duan2002}. Hence, in a co-propagation scheme, we can collect the most Stokes photons and anti-Stokes photons. 

3. The co-propagation scheme is more stable, straightforward and therefore more practical for applications. We do not need to optimize the overlap between the write pulse and the read pulse in the Cs cell as previous experiments, and one single-mode fiber can collect both Stokes and anti-Stokes photons. 

Of course, the price of these advantages is the challenge to separate strong control field, Stokes and anti-Stokes photons from a single spatial mode to three modes with high extinction ratios. By designing and constructing a filtering system, we have solved such a technical challenge nicely.

\section*{\large  Supplementary Note 4: The polarization relation between Stokes (anti-Stokes) photons and the write (read) pulse in far off-resonant Raman process }
There is no an external magnetic field applied to define the quantization axis. It can be seen that we even use a three-layer magnetic shielding to minimize the influence of the external magnetic field, which is shown in Figure\,1a. 
As there is no magnetic field applied, the choice of quantization axis can be arbitrary. For convenience, we choose the quantization axis which is vertical to the propagation direction of the write and read pulses but is parallel to the polarization of the write and read pulses. Thus, we can call the write and read pulse as a $\pi$ polarized light.

In a far off-resonance Raman transition process\cite{Nunn2008}, if the initial state is $\left| {F_{{\rm{initial}}}  = 3,m_i } \right\rangle $ and the final state is $\left| {F_{{\rm{final}}}  = 4,m_{{\rm{final}}} } \right\rangle$, the intermediate sate can only be $\left| {F'_{{\rm{intermediate}}}  = 3,m'_{{\rm{intermediate}}}  = m_i } \right\rangle $ or $
\left| {F'_{{\rm{intermediate}}}  = 4,m'_{{\rm{intermediate}}}  = m_i } \right\rangle$, as a $\pi$ polarized light cannot change the magnetic quantum number $m_i$. Thus, the final sate  $\left| {F_{{\rm{final}}}  = 4,m_{{\rm{final}}} } \right\rangle$ can be reached by two paths (the only exception is $m_i=0$ initially) from the initial state. The first path is $\left| {F_{{\rm{initial}}}  = 3,m_i } \right\rangle  \Rightarrow \left| {F'_{{\rm{intermediate}}}  = 3,m'_{{\rm{intermediate}}}  = m_i } \right\rangle  \Rightarrow \left| {F_{{\rm{final}}}  = 4,m_{{\rm{final}}} } \right\rangle$. The second path is $\left| {F_{{\rm{initial}}}  = 3,m_i } \right\rangle  \Rightarrow \left| {F'_{{\rm{intermediate}}}  = 4,m'_{{\rm{intermediate}}}  = m_i } \right\rangle  \Rightarrow \left| {F_{{\rm{final}}}  = 4,m_{{\rm{final}}} } \right\rangle$. There should be some interference between the two paths. Given a arbitrary initial state $\left| {F_{{\rm{initial}}}  = 3,m_i } \right\rangle$ ( i.e., any value of $m_i$ ), one can calculate the amplitude $A_{\pi \pi }$ for $\pi$ polarized Stokes photons excited by a $\pi$ polarized write pulse.
\begin{equation}
\begin{split}
 A_{\pi \pi } \left( {F_{{\rm{initial}}}  = 3,m_i } \right) = \sum\limits_{F'_j  = 3,4} {\left\{ {d\left( {J = \frac{1}{2},F_{{\rm{initial}}}  = 3,m_i  \to J = \frac{3}{2},F'_j ,m'_j  = m_i } \right)} \right. \times }  \\ 
\left. {d\left( {J = \frac{3}{2},F'_j ,m'_j  = m_i  \to J = \frac{1}{2},F_{{\rm{final}}}  = 4,m_{{\rm{final}}} } \right)} \right\} \\ 
\end{split}
\end{equation}
where $d$ denotes the  dipole matrix element connecting the states with angular momentum quantum numbers indicated by its arguments. If the write pulse is far detuned from the intermediate state, the energy splitting around 200\,MHz between the intermediate hyperfine levels makes a negligible difference to the coupling of the write pulse and intermediate hyperfine levels. In the far detuning condition, one can calculate out $A_{\pi\pi}$ which equals to 0, {\it i.e.}, the polarization of Stokes photons is vertical to the polarization of the write pulse. Similarly, the polarization of anti-Stokes photons is also vertical to the polarization of the read pulse. Therefore, in far off-resonant Raman process, Stokes photons and anti-Stokes photons are basically vertically polarized if the write and read pulse are horizontally polarized. This is the reason why we can use a Wollaston polarizer (WP) with an extinction ratio of 100000:1 to basically filter out Stokes and anti-Stokes photons from the write and read pulses.  

\section*{\large  Supplementary Note 5: The violation of Cauchy-Schwartz inequality}
The Cauchy-Schwarz inequality predicted classically:
\begin{equation}
\left[ {g_{\rm S - AS}^{(2)} } \right]^2  \le g_{\rm S - S}^{(2)}  \times g_{\rm AS - AS}^{(2)} 
\end{equation}
The left side should be less than or equal to the right side. The violation of Cauchy-Schwartz inequality means the left side is larger than the right side. The violation is calculated in the following form.
\begin{equation}
{\rm The\,\,violation}=\frac{{\left[ {g_{\rm S - AS}^{(2)} } \right]^2  - g_{\rm S - S}^{(2)}  \times g_{\rm AS - AS}^{(2)} }}{\rm standard{\,} deviation} 
\end{equation}
where the standard deviation $=\sqrt {\left( {{\rm{2}}g_{\rm S - AS}^{(2)}\cdot \delta g_{\rm S - AS}^{(2)} } \right)^2  + \left( {g_{\rm AS - AS}^{(2)} \cdot \delta g_{\rm S - S}^{(2)} } \right)^2  + \left( {g_{\rm S - S}^{(2)}\cdot \delta g_{\rm AS - AS}^{(2)} } \right)^2 } $, $  \delta g_{\rm S - AS}^{(2)}= g_{\rm S - AS}^{(2)}  \times \sqrt {1/N_{\rm S}  + 1/N_{\rm AS}  + 1/N_{\rm S - AS} } $, and so on. $N_{\rm S}$ is the count of Stokes photons, $N_{\rm AS}$ is the count of anti-Stokes photons, $N_{\rm S - AS}$ is the coincidence count of Stokes photons and anti-Stokes photons. The standard deviation is derived by error propagation function and Poisson distribution of photon counting.

\section*{\large  Supplementary Note 6: The interpretation of the fitting function for the storage time curve}
We express the fitting function as $g_{\rm S - AS}^{(2)}  = 1 + \frac{C}{{At^2  + Bt + 1}}$. In the ideal condition, $A = v_{\rm s}^2 /\left( {w_0^2 } \right),{\rm{ }} v_{\rm s}  = \sqrt {2k_{\rm B} T/m}$. $m$ is the mass of a Cs atom, and $w_0$ is the beam radius of the write and read light. In practice, taking account of the collision between Cs atoms and the Ne buffer gas, the diffusion speed should be much smaller than $\sqrt {2k_{\rm B} T/m}$. Using nonlinear curve fitting of Origin software, we obtain the parameters shown in Supplementary Table 2. The parameter $A$ decreases with the increase of the beam radius $w_0$, crossponding to the cross correlation $g_{\rm S - AS}^{(2)}$ increases with the increase of the beam radius. Note that the divergence angle of pulse with smaller beam waist is bigger, so the interaction region with beam waist $w_0  = 240\mu {\rm{m}}$ is about 4.9 times of the interaction region with beam waist $w_0  = 90\mu {\rm{m}}$. After the  beam waist changes from $w_0  = 90\mu {\rm{m}}$ to $w_0  = 240\mu {\rm{m}}$, the parameter $A$ should change from $1.80 \times 10^{ - 6} /{\rm{ns}}^2 $ to $\left( {1.80 \times 10^{ - 6} /{\rm{ns}}^2 } \right)/4.9  = 3.7 \times 10^{ - 7} /{\rm{ns}}^2$, which is consistent with $4.73 \times 10^{ - 7} /{\rm{ns}}^2 $ shown in Supplementary Table 2. 

The term $Bt$ is a correction that reflects the effect of the background noise during the storage time (the time between the write pulse and read pulse). In the ideal condition, this background noise should be zero. However, in practice, the leakage from Electro Optic Modulator and spontaneous emissions from the homemade tapered amplifier are not negligible. In our experiment, the switching extinction ratio of the write and read light is several hundreds, which means the write and read light is not absolutely zero when the light should be switched off. The non-zero light may destroy the stored excitation state. It is reasonable to consider the total destruction of the nonzero background noise as linear to the storage time $t$. We therefore add the correction term $Bt$ to the model \cite{Zhao2009}. The intensity of the background noise decreases with the increase of the beam waist, so the destruction coefficient $B$ should decrease with the increase of the beam waist $w_0$. As is shown in the Supplementary Table 2, $B$ decreases from $0.0013{\rm{/ns}}$ to $6.73 \times 10^{-4} {\rm{/ns}}$. So far, we have interpreted the parameters $A$ and $B$.
 
From Supplementary Table 2, one can infer that the cross correlation is mainly determined by the term $Bt$ when the storage time is much smaller than 1000\,ns, and is mainly determined by the term $At^2$ when the storage time is much longer than 1000\,ns. So we conclude that the life time of our quantum memory is mainly determined by the term $At^2$, {\it i.e.}, the average time for the atoms passing through the interaction region.
\begin{table}[h]
	\centering
     \begin{tabular}{|c|l|l|}
		\hline
		Beam waist & $90\mu {\rm{m}}$ & $240\mu {\rm{m}}$ \\ \hline
	    $A$ & $1.80 \times 10^{ - 6} /{\rm{ns}}^2 $ & $4.73 \times 10^{ - 7} /{\rm{ns}}^2 $\\ \hline
          $B$ & $0.0013{\rm{/ns}}$ & $6.73 \times 10^{ - 4} {\rm{/ns}}$ \\ \hline
          $C$ & 10.28 & 27.55\\ \hline
          Reduced Chi-Sqr & 0.10917 & 0.01572\\ \hline
          Adj. R-Square & 0.99066 & 0.99984\\ \hline
	\end{tabular}
\caption{The parameters for fitting the data in Figure\,4.}
\end{table}

\section*{\large  Supplementary Note 7: Possible solutions to prolonging the lifetime }
Possible solutions to prolong the lifetime may include: 

1. By enlarging the beam waist, the motion induced loss can be reduced and a longer lifetime has been achieved, see Figure 4 in the main text and  Supplementary Note 6. It should be noticed that the beam waist is not necessarily larger than the radius of cell, a bit close to the size of the cell will be okay, because atoms are hopping around and will pass the beam anyway if the two sizes are similar.

2. There is a spin-exchange between the atom and the wall of cell, which may annihilate the stored collective excitation. The cell in our system has no anti-relaxation coating, and the relaxation time (which is also called T1) is only tens of microseconds, which is the dominant factor for the lifetime with our current devices. The anti-relaxation coating, such as paraffin, can support approximately $10^4$ atom-wall collisions before depolarizing the atom spins \cite{Balabas2010}. With a well-coated cell, very long relaxation time can be achieved. We are collaborating with a Russian scientist who has been engaged in coating the cell.

3. In addition, reducing the background noise i.e. the leakage from Electro Optical Modulator and the spontaneous emissions from the homemade tapered amplifier is also a practical and promoted approach to prolong the lifetime.

\end{document}